\documentclass[twocolumn,showpacs,preprintnumbers,amsmath,amssymb]{revtex4}

\usepackage{graphicx}
\usepackage{dcolumn}
\usepackage{bm}

\begin{document}
\title{A length scale for the superconducting Nernst signal above T$_{c}$ in Nb$_{0.15}$Si$_{0.85}$}
\author{A. Pourret$^{1}$,  H. Aubin$^{1}$, J. Lesueur$^{1}$, C. A. Marrache-Kikuchi$^{2}$,
L. Berg\'{e}$^{2}$, L. Dumoulin$^{2}$, K. Behnia$^{1}$}
\affiliation{(1)Laboratoire de Physique Quantique(CNRS), ESPCI, 10
Rue de Vauquelin, 75231 Paris, France \\
(2)CSNSM, IN2P3-CNRS,B\^{a}timent 108, 91405 Orsay, France}
\date{January 11, 2007}

\begin{abstract}
We present a study of the Nernst effect in amorphous superconducting
thin films of Nb$_{0.15}$Si$_{0.85}$. The field dependence of the
Nernst coefficient above T$_{c}$ displays two distinct regimes
separated by a field scale set by the Ginzburg-Landau correlation
length. A single function $F(\xi)$, with the correlation length as
its unique argument set either by the zero-field correlation length
(in the low magnetic field limit) or by the magnetic length (in the
opposite limit), describes the Nernst coefficient. We conclude that
the Nernst signal observed on a wide temperature ($30 \times T_c$)
and field ($4 \times B_{c2}$) range is exclusively generated by
short-lived Cooper pairs.
\end{abstract}


\maketitle

The observation of a finite Nernst signal in the normal state of
high-$T_c$ cuprates\cite{Wang2006} has revived interest in the study
of superconducting fluctuations. In low-temperature conventional
superconductors, short-lived Cooper pairs above $T_c$ have been
mostly examined through the phenomena of
paraconductivity\cite{Glover1967} and fluctuation
diamagnetism\cite{Gollub1973}. Due to a sizeable contribution coming
from free electrons to conductivity and magnetic susceptibility, the
sensitivity of these probes to superconducting fluctuations is
limited to a narrow region close to the superconducting
transition\cite{Skocpol1975}.

Because of a low superfluid density, the superconducting state in
underdoped cuprates is particularly vulnerable to phase
fluctuations\cite{emery95}. Therefore, long-lived Cooper pairs
without phase coherence and vortex-like excitations associated with
them were considered as the main source of the anomalous Nernst
effect observed in an extended temperature window above T$_c$ in
underdoped cuprates\cite{Wang2006}.

In a recent experiment on amorphous thin films of the conventional
superconductor Nb$_{0.15}$Si$_{0.85}$\cite{Pourret2006}, we found
that a Nernst signal generated by short-lived Cooper pairs  could be
detected up to very high temperatures ($30 \times T_c$) and high
magnetic field ($4 \times B_{c2}$) in the normal state. In these
amorphous films, the contribution of free electrons to the Nernst
signal is negligible. Indeed, the Nernst coefficient of a metal
scales with electron mobility\cite{behnia06}. The extremely short
mean free path of electrons in amorphous Nb$_{0.15}$Si$_{0.85}$
damps the normal-state Nernst effect and allows a direct comparison
of the data with theory. In the zero-field limit and close to $T_c$,
the magnitude of the Nernst coefficient was found to be in
quantitative agreement with a theoretical
prediction\cite{Ussishkin2002} by Ussishkin, Sondhi and Huse, (USH)
invoking the superconducting correlation length as its single
parameter. At high temperature and finite magnetic field, the data
was found to deviate from the theoretical expression.

\begin{figure}
{\includegraphics[scale=0.7]{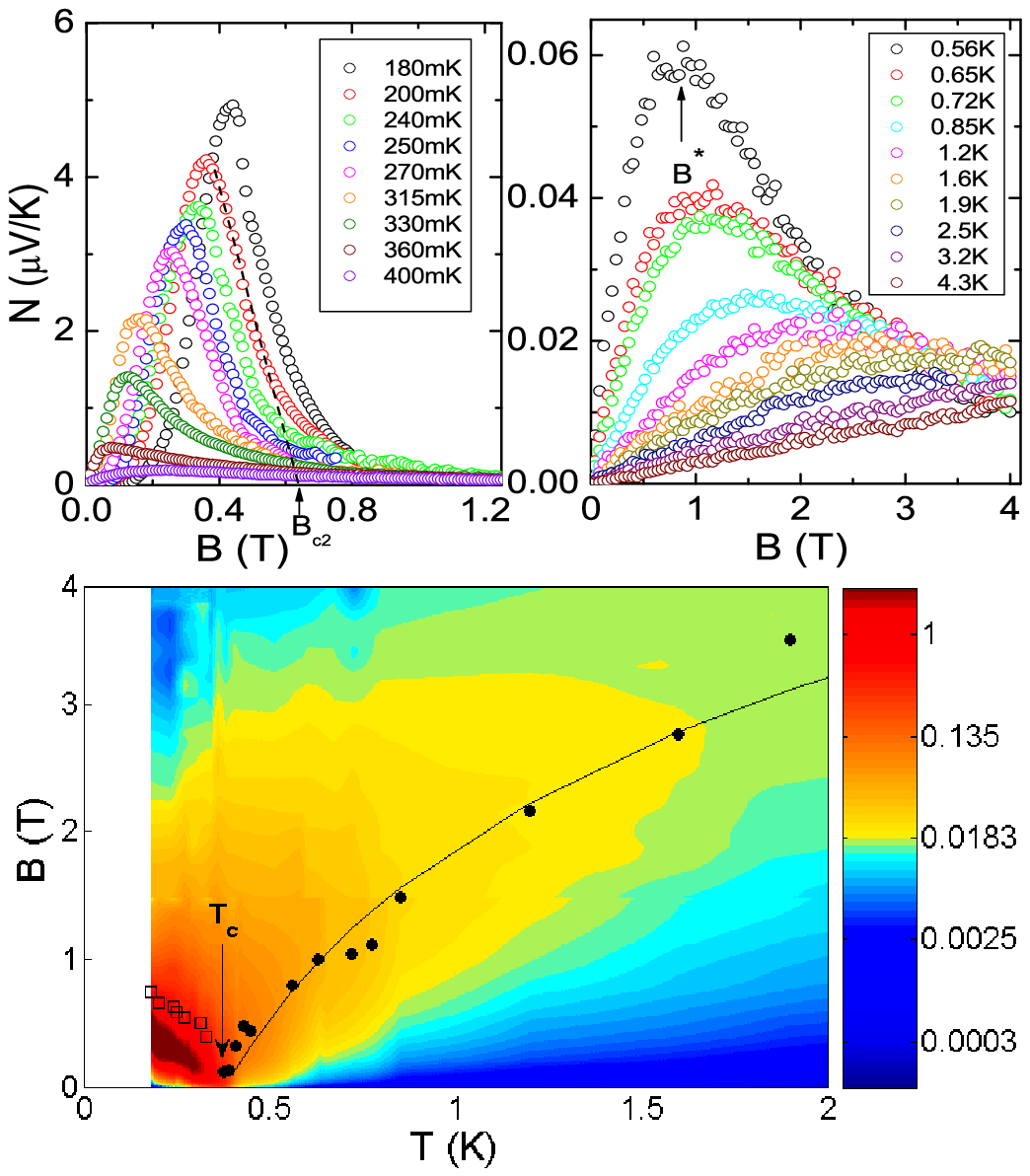}} \caption{\label{fig:fig1}
The Nernst signal (N) as a function of magnetic field for
temperatures ranging from 0.180 K to 0.360 K (Upper left panel) and
from 0.56 to 4.3 K (Upper right panel) measured on sample 2
($T_c$=380 mK). Above (below) $T_c$ the position of the maximum in
the field dependence of the Nernst signal increases (decreases) with
decreasing temperature. Lower panel : Logarithmic color map of the
Nernst signal in the (B,T) plane. Superimposed on the plot are the
values of the critical field (open squares), below $T_c$, and $B^*$
( full circles), above $T_c$. Note the symmetric evolution of these
two with respect to $T_c$. The continuous line is the field scale
set by the Ginzburg-Landau correlation length,
$B^{*}=\frac{\phi_{0}}{2\pi \xi_{d}^2}$. }
\end{figure}

In this Letter, we extend our measurements and analysis of the
Nernst signal to high magnetic field and temperatures well above
T$_{c}$. The Nernst coefficient, $\nu$, is reduced as one increases
either the temperature or the magnetic field. We will show here that
both these variations reflect a unique dependence on a single length
scale. A striking visualization of this emerges when one substitutes
temperature and magnetic field by their associated length scales:
the zero-field superconducting correlation length $\xi_d(T)$ and the
magnetic length $l_B(B)={(\hbar/2eB)}^{1/2}$. The symmetric contour
lines of $\nu$ in the ($l_B$, $\xi_d$) plane shows that its
dependence on both field and temperature can be described by a
single function with the superconducting correlation length $\xi$ as
its unique argument. In the low-field limit, the correlation length
is set by $\xi_d$ and in the high-field limit by $l_B$. In the
intermediate regime, when $\xi_d \simeq l_B$, the correlation length
is a simple combination of these two lengths. This observation is
additional proof that the Nernst signal observed up to high
temperature ($30\times T_c$) and high magnetic field ($4 \times
B_{c2}$) in this system is exclusively generated by superconducting
fluctuations -- i.e. short-lived Cooper pairs. Hence the functional
dependence of the Nernst coefficient on the correlation length is
empirically determined in a wide range extending from the long
correlation length regime, where the data follows the prediction of
USH theory, to the short correlation length regime, where the
Ginzburg-Landau approximation fails and no theoretical expression is
yet available.

Amorphous thin films of Nb$_x$Si$_{1-x}$ were prepared in ultrahigh
vacuum by electron beam co-evaporation of Nb and Si, with special
care over the control and homogeneity of the concentrations. The
competition between superconducting, metallic and insulating ground
states is controlled by the Nb concentration, the thickness of the
films, or the magnetic field\cite{Marnieros2000, Lee2000,
Aubin2006}. Two samples of identical stoichiometry --
Nb$_{0.15}$Si$_{0.85}$ -- but with different thicknesses and $T_c$
are used in this study. Sample 1 (2) was 12.5 (35) nm thick  and its
midpoint $T_c$ was 0.165 (0.380) K.  The physical properties of
these two samples as well as the experimental set-up were detailed
in our previous communication, which also presented part of the data
discussed here\cite{Pourret2006}.

Figure~\ref{fig:fig1} shows the Nernst signal,
$N=\frac{E_{y}}{(-\nabla_{x}T)}$ as a function of magnetic field and
temperature, for sample 2. Below $T_{c}$, the magnetic field
dependence shows the characteristic features of vortex-induced
Nernst effect, well known from previous studies on conventional
superconductors\cite{Huebener1969} and high-$T_c$
cuprates\cite{Wang2006}. For each temperature, the Nernst signal
increases steeply when the vortices become mobile following the
melting of the vortex solid state. It reaches a maximum and
decreases at higher fields when the excess entropy of the vortex
core is reduced. As the temperature decreases the position of this
maximum shifts towards higher magnetic fields. This is not
surprising, since in the superconducting state, all field scales
associated with superconductivity are expected to increase with
decreasing temperature. In particular, this is the case of the upper
critical field, $B_{c2}$, which can be roughly estimated by a linear
extrapolation of the Nernst signal to zero as it has been done in
cuprates\cite{Wang2006}.

Above $T_c$, the temperature dependence of the characteristic field
scale is reversed. In this regime, at low magnetic field, the Nernst
signal $N=R_{square}\times \alpha_{xy}$ increases linearly with
field, as expected from the USH theory where $\alpha_{xy}$ follows
the simple expression\cite{Ussishkin2002} :
\begin{equation}
\alpha_{xy}=\frac{1}{3\pi}\frac{k_Be}{\hbar}\frac{\xi^2} {{l_B}^2}
\end{equation}

[Note that here we use the definition of
$l_B(B)={(\hbar/2eB)}^{1/2}$, which differs by a factor of
$\sqrt{2}$ from what was used in ref.\cite{Ussishkin2002} and
\cite{Pourret2006}. This, for obvious reasons to appear below.] Upon
increasing the magnetic field, the Nernst signal deviates from this
linear field dependence, reaches a maximum at a field scale $B^*$
and decreases afterwards to a weakly temperature-dependent
magnitude. In contrast to the superconducting state ($T<T_c$), the
position of the maximum shifts to higher fields with increasing
temperature. The contour plot in the (T,B) plane (lower panel) shows
that these two field scales, $B_{c2}$ and $B^*$, evolve
symmetrically with respect to the critical temperature. One major
observation here is that the magnitude and the temperature
dependence of $B^*$ follows the field scale set by the
Ginzburg-Landau correlation length
$\xi_{d}=\frac{\xi_{0d}}{\sqrt{\epsilon}}$ through the relation
$B^{*}=\frac{\phi_{0}}{2\pi \xi_{d}^2}$ where $\phi_{0}$ is the flux
quantum and $\epsilon=\ln{\frac{T}{T_{c}}}$ the reduced temperature.
[This field scale  was dubbed ``the ghost critical field'' by
Kapitulnik and co-workers\cite{Kapitulnik1985}.] The value for the
correlation length at zero-temperature $\xi_{0d}$ is determined from
the BCS formula in the dirty limit $\xi_{0d}= 0.36
\sqrt{\frac{3}{2}\frac{\hbar v _{F}\ell}{k_{B}T_{c}}}$, where
v$_{F}$ $\ell = 4.35\times 10^{-5}$ m$^{2}$s$^{-1}$ is estimated
using the known values of electrical conductivity and specific
heat\cite{Pourret2006}.
\begin{figure}[h!]
{\includegraphics[scale=0.8]{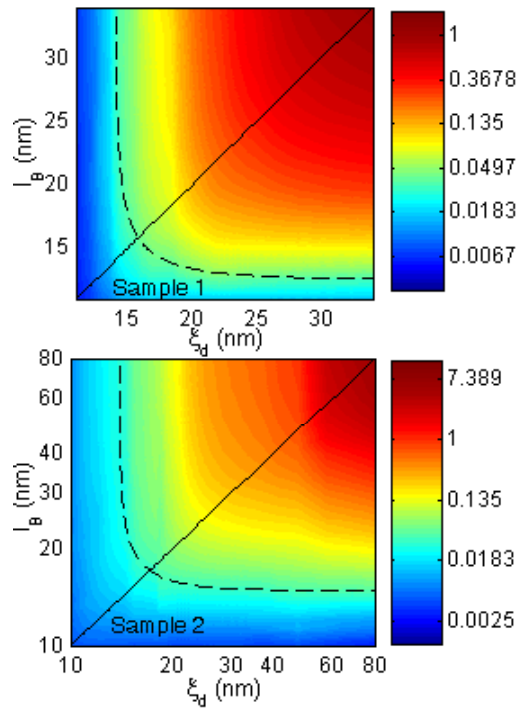}} \caption{\label{fig:fig2}
Logarithmic color map of the Nernst coefficient as a function of the
magnetic length $l_B$ and the zero-field correlation length $\xi_d$,
for sample 1 (upper panel) and sample 2 (lower panel). Note the
symmetry of the Nernst coefficient with respect to the diagonal
continuous line ($l_{B}=\xi$). Dots lines represents contours for
$\xi=15 nm$ with $\xi=(1/{\xi_d}^4+1/{(c\times l_B)}^4)^{-1/4}$.
$c=1.12$ for sample 1 and $c=0.93$ for sample 2.}
\end{figure}
Thus, the decrease of the Nernst signal at the field scale $B^*$ is
the consequence of the reduction of the correlation length when the
cyclotron diameter $l_B$ becomes shorter than the zero-field
correlation length $\xi_{d}$, at a given temperature. This phenomena
is well known from studies of fluctuations diamagnetism in low
temperature superconductors\cite{Gollub1973} and
cuprates\cite{Carballeira2000}. While in the low field limit, the
magnetic susceptibility should be independent of the magnetic field
-- i.e. in the Schmidt limit\cite{Schmid1969} --, the magnetic
susceptibility is experimentally observed to decrease with the
magnetic field, following the Prange's formula\cite{Prange1970};
which is an exact result within the Ginzburg-Landau formalism that
takes into account the reduction of the correlation length by the
magnetic field. In this regime, the amplitude fluctuations are
described as evanescent Cooper pairs arising from free electrons
with quantized cyclotron orbits\cite{Skocpol1975}. In our
experiment, we can clearly distinguish the low-field limit -- where
the cyclotron length is larger than the correlation length -- from
the high-field limit, where the correlation length has been reduced
from its value at zero field to a shorter value given by the
magnetic length.

Figure~\ref{fig:fig2} presents a contour plot (with a logarithmic
scale of the colors) of the Nernst coefficient $\nu=N/B$ in the
($l_{B}$, $\xi_{d}$) plane. [Note that in contrast to
Figure~\ref{fig:fig1}, $\nu$ and not $N$ is plotted here]. These
graphs are instructive. When $l_{B}>\xi_{d}$, the contour lines are
parallel to the magnetic length axis, meaning that the Nernst
coefficient depends only on $\xi_{d}$. On the other side of the
diagonal, when $l_{B}<\xi_{d}$, the contour lines are parallel to
the correlation length axis, meaning that the Nernst coefficient
depends only on $l_B$. Furthermore, in both samples, the contour
lines are almost symmetric with respect to the $l_{B}=\xi_{d}$ line.
In other words, the Nernst coefficient appears to be uniquely
determined by the correlation length, no matter whether $\xi_d$ or
$l_B$ sets it.

\begin{figure}[h!]
{\includegraphics[scale=0.85]{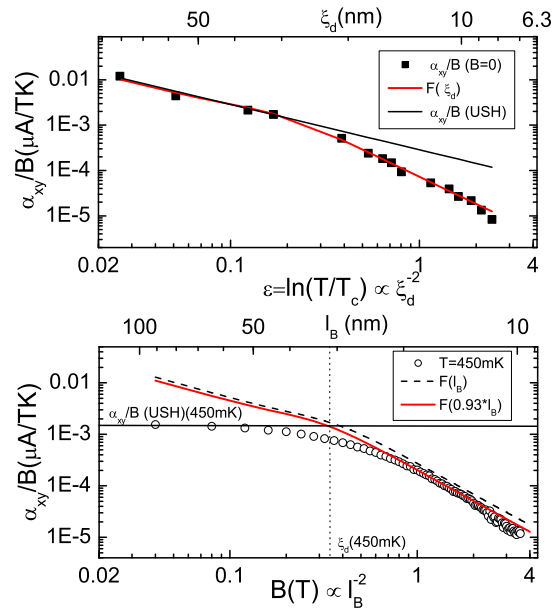}}
\caption{\label{fig:fig3}Upper panel: The transverse Peltier
coefficient $\alpha_{xy}$ divided by the magnetic field B (black
squares) as a function of the reduced temperature $\epsilon$ (bottom
axis) and correlation length (top axis) for sample 2 on a log-log
scale. At low $\epsilon$, the experimental data are consistent with
the USH model(solid line). The function $F(\xi)$ shown as a red line
is a fit to the data. Lower panel : The same function, (dot-line)
plotted using $l_B$ as its argument, $F(l_B)$, as a function of the
magnetic field (bottom axis) and $l_B$ (top axis). Note that it is
consistent with the magnitude and field dependence of the data
acquired at high magnetic field. The function $F(c\times l_B)$
(continuous line) describes nicely the data when $c=0.93$. At low
magnetic field, when $l_B>\xi_d$, the data deviates from $F(c\times
l_B)$ to reach the field-independent value $F(\xi_d)$.}
\end{figure}

Therefore, we are entitled to assume that there exists a function
which links $\nu$ to the correlation length. In the zero-field
limit, the latter is set by $\xi_{d}$, and we can empirically
extract from our data a function $F(\xi)$ such as
$F(\xi)=\alpha_{xy}/B=\frac{\nu}{R_{square}}(T,B\rightarrow 0)$.
This function is shown in figure~\ref{fig:fig3}(upper panel). For
large $\xi$, it is proportional to $\xi^2$, in agreement with the
USH expression. For small $\xi$, however, we find that this function
becomes roughly proportional to $\xi^4$. Now a remarkable
observation follows : when this function $F(\xi)$ is plotted with
the magnetic length $l_B$ as argument, as shown in the lower panel
of figure ~\ref{fig:fig3}, we find that it almost describes the
magnitude and the field dependence of the Nernst coefficient at high
magnetic field. This demonstrate that in these two opposite limits,
($\xi\simeq\xi_d$ and $\xi\simeq l_B$ ), the Nernst coefficient is
determined by this single function $F(\xi)$ that depends uniquely on
the correlation length.

\begin{figure}[h!]
{\includegraphics[scale=1]{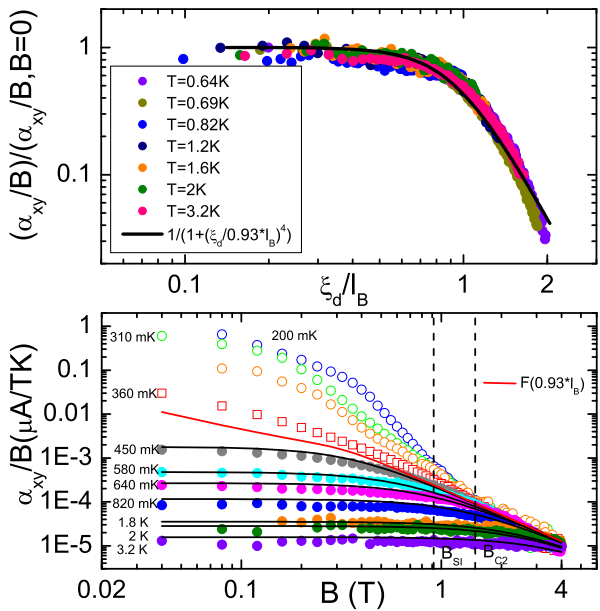}} \caption{\label{fig:fig4}
Upper Panel: $\alpha_{xy}/B$  normalized to its zero-field value
versus $\xi_{d}/l_{b}$ for sample 2. All the data at temperatures
between 430 mK and 3.2 K are shown here to collapse on
$\frac{1}{{1+{(\xi_d/0.93*l_B)}^4}}$ (thick line). Lower panel:
$\alpha_{xy}/B$ as a function of $B$ for temperatures ranging from
200 mK to 3.2 K on a log-log scale. The function
$F(\xi)=\frac{\alpha_{xy}}{B}(\xi_d)$ extracted from the data at
zero field is plotted here as a function of $l_B$ (thick red line).
At high magnetic field, all the data, above and below $T_c$ tend
towards $F(c\times l_B)$. Note that $F(c\times l_B)$ is the
separatrix of the data above and below $T_c$. Also shown is $F(\xi)$
for several values of $\xi_d$ corresponding to different
temperatures, and plotted as function of
$\xi=(1/\xi_{d}^{4}+1/(0.93*l_{B})^4)^{-1/4}$.}
\end{figure}

This function, extracted from the zero field data and observed to
describe the data at high magnetic field, should also be valid at
intermediate field, when $\xi_d\simeq l_B$. In this intermediate
region, with the increasing magnetic field, the correlation length
progressively evolves from $\xi_{d}$ to $l_B$. A simple relation for
$\xi$ that verifies the conditions, $\xi\simeq\xi_{d}$ when
$l_B\rightarrow \infty$, and $\xi\simeq c\times l_B$ when
$l_B\rightarrow 0$ is given by :
\begin{equation} \label{eq:eq1}
\frac{1}{\xi^{\gamma}}=\frac{1}{{\xi_d}^{\gamma}}+\frac{1}{(c\times
l_{B})^{\gamma}}
\end{equation}
where the pre-factor c and the exponent $\gamma$ are to be
determined from the experimental data. The pre-factors $c=1.12$ for
sample 1 and $c=0.93$ for sample 2 are determined such as $F(c\times
l_B)$ gives precisely the field dependence of the Nernst coefficient
at high magnetic field, shown figure \ref{fig:fig3}. The slight
difference between the pre-factors for the two samples is not
understood at this stage. It appears to be larger than the
experimental margin of error and suggests that other parameters not
taken into account in this analysis may be involved, such as the
thickness of the samples. To get the value of the exponent $\gamma$,
we first note that the curves
$\frac{\alpha_{xy}/B(B,T)}{\alpha_{xy}/B(B\rightarrow 0,T)}$, for
temperature $T>2\times T_c$, collapse on a unique curve when they
are plotted as a function of $\xi_d/l_B$, as shown figure
\ref{fig:fig4}. Then, using $F(\xi)\sim\xi^4$ -- as previously
observed for $T>2\times T_c$ -- and using equation~\ref{eq:eq1}, we
find that $F(\xi)/F(\xi_d)=[1+{(\frac{\xi_d}{c\times
l_B})}^{\gamma}]^{-4/\gamma}$ depends only on the ratio $\xi_d/l_B$,
and so has the appropriate functional form to describe the collapsed
data. For both samples, the best fit is obtained when $\gamma=4$, as
shown Figure \ref{fig:fig4} for sample 2. Such a value of $\gamma$
implies that the first non-linear correction to the field dependence
of $\alpha_{xy}$ is proportional to $B^2$(i.e $l_B^{-4}$), in
agreement with analyticity arguments\footnote{M. Feigelman, Private
Communication}. With these parameters just determined, figure
\ref{fig:fig2} shows that the contour lines of equal Nernst
coefficient can be described by curves of constant correlation
length $\xi$ as given by equation \ref{eq:eq1}. Thus, as shown
figure~\ref{fig:fig4}, above $T_{c}$, the magnitude of the Nernst
coefficient at any temperature and magnetic field is given by this
unique function $F(\xi)=\frac{\alpha_{xy}}{B}(\xi)$, determined
experimentally at zero field, where the correlation length is given
by equation \ref{eq:eq1}.

Figure ~\ref{fig:fig4} also shows that the magnitude and field
dependence of the Nernst signal, measured for temperatures $T<T_c$
and high magnetic field ($B>B_{c2}$), can also be described by
$F(c\times l_B)$. In particular, we see that the Nernst data
measured at a temperature close to $T_c$ follows closely the curve
$F(c\times l_B)$ on a wide magnetic field range; this is expected
for the diverging Ginzburg-Landau correlation length when
approaching $T_c$. It is remarkable to note that the curve $F(\xi)$,
extracted from the data at zero-field, and plotted as a function of
$l_B$, gives precisely the separatrix between the curves
$\frac{\alpha_{xy}}{B}(B)$ measured above and below $T_c$. Below
$T_c$, $\alpha_{xy}/B$ joins the function $F(c\times l_B)$ at a
critical field $B_{c2}\simeq 1.5~T$, which is larger than the
critical field of the superconductor-insulator transition,
$B_{SI}=0.91~T$, defined as the crossing field of $R(B)$
curves\cite{Aubin2006}.

The overall consistency of this analysis demonstrates that the
off-diagonal component of the Peltier tensor, $\alpha_{xy}$, is set
by $\xi_d$ over a wide temperature range. As noticed previously,
when $\xi_d$ is large, $\alpha_{xy}$ is consistent with USH
expression. However, below a correlation length of the order of
$\xi_{0d}$, it deviates downward from the USH formula. Such a
deviation is actually expected for theories based on the
Ginzburg-Landau formalism, which are notoriously known to
overestimate the effect of short-wave length
fluctuations\cite{Skocpol1975}. In fluctuations diamagnetism
experiments, the data were observed to fall systematically below the
Prange's limit above some scaling field, of the order of $0.5\times
H_{c2}$ in dirty materials, i.e. $\xi\simeq\xi_{0d}$.

Let us conclude by  briefly commenting the case of cuprates. One
crucial issue to address is the existence of the ``ghost critical
field'' detected here. We note that such a field scale has not been
identified in the analysis of the cuprate data\cite{Wang2006}.
However, one shall not forget that the normal state of the
underdoped cuprates is \emph{not} a simple dirty metal. The
contribution of normal electrons to the Nernst signal is too large
to be entirely negligible. This makes any comparison between theory
and data less straightforward than the analysis performed here.
Moreover, in Nb$_{0.15}$Si$_{0.85}$ there is a single temperature
scale, $T_{BCS}$, for the destruction of superconductivity. In
contrast, the situation in the cuprates may be complicated by the
possible existence of two distinct temperature scales, the pairing
temperature and a Kosterlitz-Thouless-like temperature leading to a
phase-fluctuating superconductor\cite{podolsky06}.

To summarize, in the regime of short-lived Cooper pairs, the Nernst
coefficient was found to only depend on the superconducting
correlation length. We could clearly distinguish the regime where
the correlation length is set by the Ginzburg-Landau correlation
length at zero-field from the high-field regime where the
correlation length is set by the magnetic length. In the
intermediate region where $\xi_d \simeq l_B$, the correlation length
is a simple combination of these two lengths. These results
demonstrate that the Nernst signal observed at high field and
temperature in amorphous films of Nb$_{0.15}$Si$_{0.85}$ is
generated by superconducting fluctuations characterized by
short-lived Cooper pairs, implying both amplitude and phase
fluctuations of the superconducting order parameter.

We acknowledge useful discussions with S. Caprara, M.V. Feigelman,
M. Grilli, D. Huse, S. Sondhi , I. Ussishkin and A.A. Varlamov. This
work was partially supported by Agence Nationale de la Recherche and
Fondation Langlois through ``Prix de la Recherche Jean Langlois''.

----------------------------------
\bibliography{Super_Insulator1}

\end{document}